# All-Optical Generation and Steering of Spatial Solitons in Discrete Waveguide Array

**Clark A. Merchant, Igor V. Mel'nikov,[1] and J. Stewart Aitchison**

*The Edward S. Rogers Sr. Department of Electrical and Computer Engineering, University of Toronto*
*10 King's College Road, Toronto, Ontario, Canada M5S 3G4*

**Abstract:** We numerically study Stokes beam generation in an array of discrete waveguides. The feasibility of generation and steering of spatial solitons due to nonlinear Raman response is demonstrated for the first time.

## 1. Introduction

In the past two decades there has been a growing interest in the study of discrete spatial solitons. One of the most important effects seen in discrete soliton propagation is anomalous diffraction, which has been used to demonstrate effects such as blocking, routing, time gating and other logic functions [1,2]. Discrete AlGaAs waveguides have been used to explore these nonlinear effects [1,3], but the effect of two photon absorption at higher powers prevents this material from being practical for these discrete waveguide devices. There are two ways to overcome this problem, namely either introduce gain to the system (which seems impractical), or to find another medium with good nonlinear properties, but without the problems inherent with AlGaAs.

The use of highly efficient Raman materials such as potassium gadolinium tungstate $KGd(WO_4)_2$ (KGW) offers potential for use as discrete waveguide arrays. Its high nonlinear coefficient of $n_2 \sim 36 \times 10^{-16}$ $cm^2/W$ [3] and efficient Raman shifting properties, allow for highly nonlinear devices with outputs in the telecommunications spectrum without the problems inherent to AlGaAs. Recent works on creating KGW planar waveguides via ion implantation [4] show promise towards the use of this technique in forming discrete waveguide arrays. This report explores some the all-optical generation and beam steering effects that can be possible with a highly Raman efficient material such as KGW in a discrete waveguide array.

## 2. Theory

The behavior of the input and the stimulated Raman Scattering (SRS) beams are governed in the waveguide array as shown in equations (1) as normalized functions:

$$i\frac{\partial U_j}{\partial z} + c_u\left(U_{j+1} + U_{j-1}\right) = -iQ_j V_j - g_U \left|U_j\right|^2 U_j,$$

$$i\frac{\partial V_j}{\partial z} + c_v\left(V_{j+1} + V_{j-1}\right) = -iQ_j^* U_j - g_V \left|V_j\right|^2 V_j, \qquad (1)$$

$$\frac{\partial Q_j}{\partial t} = -(\gamma - i\Delta)Q_j + U_j V_j^*,$$

---

[1] *Current address: Comtex Consulting Inc., 162 Princess Margaret Blvd., Etobocoke, Ontario, Canada M9B 2Z5; he can be reached at Tel: (416)946-8663 Fax: (416)971-3020 Email: igor.melnikov@utoronto.ca*

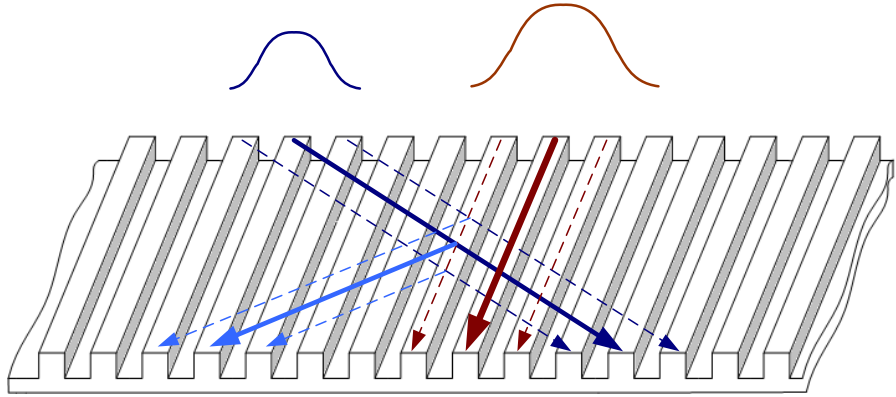

Fig. 1. Illustration showing the fundamental (blocker) beam located in the center of the waveguide array, and a Raman signal beam, illustrating transmission and reflection of the signal, with transmission occurring with no blocking beam present.

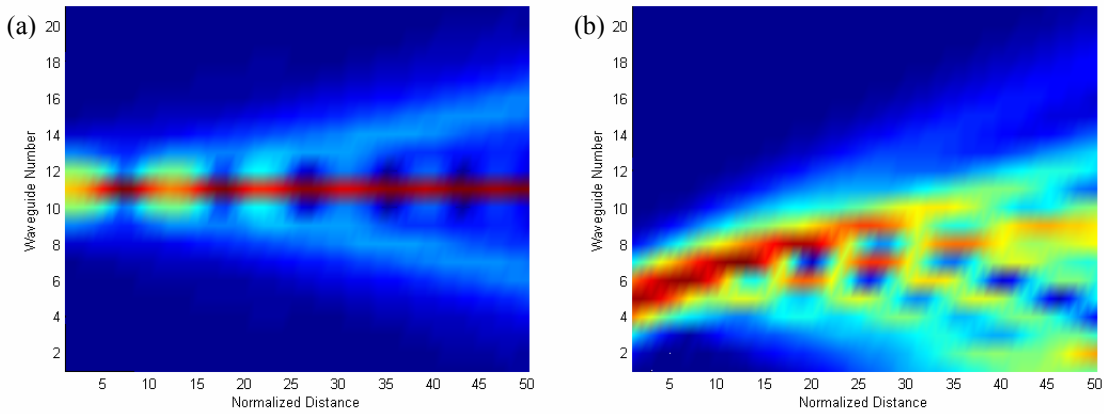

Fig. 2. Simulation of the independent propagation of the (a) input blocker beam at waveguide array center and (b) Raman signal beam applied several waveguides below the center, with an input tilt. The signal beam illustrates the expected nonlinear discrete diffraction.

Here the input signal is given as $U(z)$, and the Raman beam is designated as $V(z)$; the waveguide coupling coefficients for the input and the Raman signals are given as $c_{u,v}$, the nonlinear coefficient given as $g$, $Q$ is the effective Raman macroscopic polarization, $\gamma$ is a phenomenological dephasing coefficient, $\Delta$ is the detuning from the Raman resonance which is assumed to be negligible, and subscript $j$ gives the waveguide number in the array.

For modeling this system, the time derivative was taken as a constant. The input of the array is given a Gaussian intensity distribution across the center waveguides as illustrated in Fig. 1. The Raman beam is input several waveguides from the center of the array, and a linear phase $q$ is applied across each waveguide input to simulate a beam tilt. The adjustment of the input intensity of the blocking beam can allow for different degrees of beam steering.

3. Simulation

First the independent propagation both of the blocker and the signal beam were numerically simulated and shown in Fig. 2. The input signal was modeled with an intensity of 0.75 times the blocker input intensity, and with an angle $q=\pi/2$ for the beam tilt. Note that the tilted Raman beam first begins to channel, and then begins to illustrates nonlinear discrete diffraction as it propagates through the system.

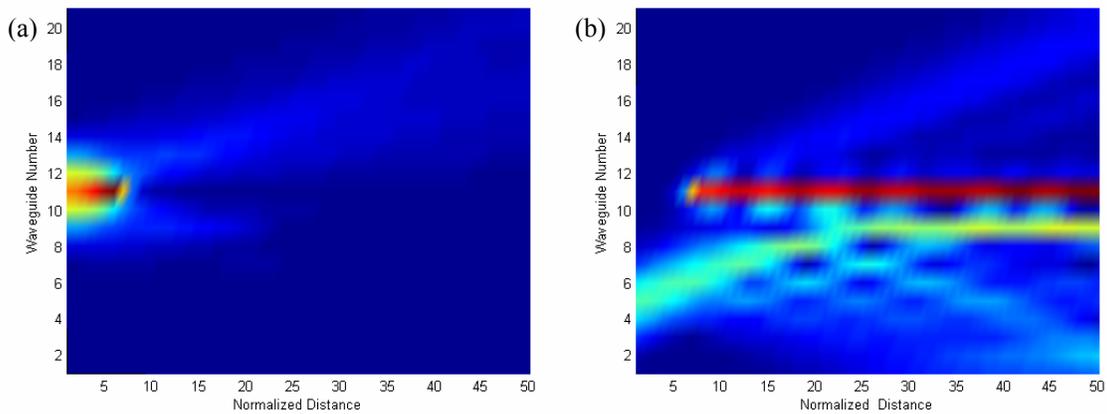

Fig. 3. Simulation of energy transfer and beam steering. The blocking beam in a) disappears, generating the SRS beam shown in b), and the subsequent beam steering of the input signal.

When the input signal propagates with an initial angle in the presence of the blocking beam, there are two effects. Firstly, there is an energy transfer from the blocking beam to the Raman wavelength which continues to propagate in the same waveguide as the fundamental blocker beam. Secondly, a steering effect is seen on the signal beam as governed by the coupled differential equations (1). As the input Raman signal couples with the blocking beam it hastens the SRS generation between the fundamental blocking wavelength and the Raman wavelength. This energy transfer and subsequent disappearance of the blocker signal at the fundamental wavelength is illustrated in Fig. 3 (a), and the angled Raman signal is seen to be channeled parallel to the stimulated Raman signal. By varying the intensity of the blocking beam, steering of the Raman signal to a different waveguide outputs can be performed.

## 4. Conclusion

We have demonstrated using numerical methods the possibility of beam steering of spatial solitons in a discrete waveguide array using Raman processes. With the availability of new highly nonlinear and Raman efficient materials such as KGW, as well as new developments in creating waveguides with these materials, there is much promise shown for applications of all optical switching, blocking, and logic functionality. Future work on discrete soliton blocking and steering can be directed towards examination of new features that may be brought by introducing non-stationary Raman response and additional phase mismatch in the waveguides.